\documentclass[12pt, a4paper]{article}
\pdfoutput=1
\usepackage{graphicx}
\usepackage{amssymb}
\usepackage{amsmath}
\usepackage{bm}
\usepackage{color}
\usepackage{cite}
\usepackage{subfigure}
\usepackage{epstopdf}            
\usepackage{epsfig}

\newcommand{\beq}{\begin{equation}}
\newcommand{\eeq}{\end{equation}}
\newcommand{\ov}{\overline}

\newcommand{\Slash}[1]{{\ooalign{\hfil/\hfil\crcr$#1$}}}

\usepackage[colorlinks=true, linkcolor=black, citecolor=black,
urlcolor=black]{hyperref}

\begin{document}

\begin{titlepage}

\begin{flushright}

IPMU15-0083 \\
FTPI-MINN-15/31

\end{flushright}

\vskip 2cm
\begin{center}

{\Large
{\bf 
Interpretations of the ATLAS Diboson Resonances
}
}

\vskip 2cm

Junji Hisano$^{1,2,3}$,
Natsumi Nagata$^{3,4}$,
and 
Yuji Omura$^{1}$

\vskip 0.5cm

{\it $^1$
Kobayashi-Maskawa Institute for the Origin of Particles and the
Universe, \\ Nagoya University, Nagoya 464-8602, Japan}\\[3pt]
{\it $^2$Department of Physics,
Nagoya University, Nagoya 464-8602, Japan}\\[3pt]
{\it $^{3}$
Kavli IPMU (WPI), UTIAS, The University of Tokyo, Kashiwa, Chiba
 277-8583, Japan}\\ [3pt] 
{\it $^{4}$William I. Fine Theoretical Physics Institute, School of
 Physics and Astronomy, University of Minnesota, Minneapolis, MN 55455,
 USA}

\vskip 1.5cm

\begin{abstract}

 The ATLAS collaboration has reported excesses in searches for resonant
 diboson production decaying into hadronic final states. This deviation
 from the Standard Model prediction may be a signature of an extra
 bosonic particle having a mass of around $2$~TeV with a fairly narrow
 width, which implies the presence of a new perturbative theory at the
 TeV scale. In this paper, we study interpretations of the signal and
 its implication to physics beyond the Standard Model. We find that the
 resonance could be regarded as a leptophobic vector particle, which
 could explain a part of the observed excesses without conflict with
 the present constraints from other direct searches for heavy
 vector bosons at the LHC as well as the electroweak precision
 measurements.

\end{abstract}

\end{center}
\end{titlepage}

\section{Introduction}
\label{sec:intro}

Recently, the ATLAS collaboration reports excesses in searches for
massive resonances decaying into a pair of weak gauge bosons
\cite{Aad:2015owa}. These excess events have been observed in the hadronic
final states, {\it i.e.}, the $pp \to V_1V_2 \to 4j$ ($V_{1,2}=W^\pm$ or
$Z$) channels. The weak gauge bosons from the resonance are
highly boosted so that the hadronic decay products are
reconstructed as two fat jets. Constructing the invariant
mass of these two fat jets, it is possible to find a resonant peak for
the intermediate state. The ATLAS collaboration has performed such an
analysis by using 20.3~fb$^{-1}$ data of the 8~TeV LHC running. Then,
the excesses with narrow widths are observed around $2$~TeV in the $WZ$,
$WW$, and $ZZ$ channels with local significance of 3.4$\sigma$,
2.6$\sigma$, and 2.9$\sigma$, respectively. Although we should wait for forthcoming ATLAS/CMS
results of relevant searches to
obtain a robust consequence about the observation, it should be worthwhile to
consider possible interpretations of these anomalous events as
evidence for new physics beyond the Standard Model (SM). 
In fact, the excesses are well fitted with resonances whose peaks are 
around 2~TeV and widths are less than about 100~GeV. Such narrow
resonances may imply new weakly-interacting particles, and then 
the underlying theories would be perturbative.\footnote{A possibility of strong dynamics is proposed and investigated in Ref.~\cite{Fukano:2015hga}.} In this paper, we especially 
consider such a possibility to explain the excesses.

As mentioned above, the excesses reported by the ATLAS collaboration are in
the $WZ$, $WW$, and $ZZ$ channels. The tagging selections for each mode
used in the analysis are, however, rather incomplete: 
about 20\% of the events are shared by these channels. 
At the present stage, it may be hard
to conclude that one resonance is responsible for the excesses
in all the channels.
There may be a possibility that one 2-TeV particle contributes to only one part of the channels 
and the peaks in the other channels are merely contamination due to the incomplete tagging
selections. Taking this situation into account, in this paper, we do not
limit ourselves to account for all of these excesses simultaneously, and
consider the possibility that the new resonance appears in one channel.
For each channel,
the number of excess events could be accounted for if there is a 2~TeV
resonance whose production cross section times decay branching ratio
into gauge bosons is about 6 fb. 
We regard this as a reference value in what follows.

In order for a resonance to decay into two gauge bosons, it should be a
bosonic state, namely, a particle with a spin zero or one under an assumption of the renormalizable theory. Let us
first consider the spin-zero case. If such a particle is a singlet under
the $\text{SU}(2)_L \otimes \text{U}(1)_Y$ gauge interactions, it
couples to the electroweak gauge bosons and the SM fermions only through
the mixing with the SM Higgs boson in the renormalizable potential. Therefore, its production cross
section is suppressed by the mixing factor and in general too small to
explain the anomalies. A possible way to enhance the production is to
introduce new vector-like colored particles. A singlet scalar field
generically couples to these colored particles. Then the
singlet is produced via the gluon fusion process according to the loop correction involving
the vector-like colored particles. If the masses
of the vector-like particles are above 1~TeV, such a scalar resonance
with 2 TeV mass does not decay into these particles. It turns out,
however, that ${\cal O}(10)$~fb production cross sections require
${\cal O}(10)$ extra colored particle pairs. Moreover, 
a large fraction of produced singlets decays into gluons, and
thus gives only a negligible contribution to the diboson channels. Hence,
a singlet scalar boson is inappropriate to explain the anomalies. An
alternative possibility is to exploit $\text{SU}(2)_L$ doublet scalars. 
These scalar fields may develop a finite vacuum expectation value (VEV)
to directly couple to $W$ and $Z$, and again mix with the SM Higgs
field. To assure a large cross section to the diboson decay processes,
there should be a sizable deviation from the SM limit. 
In addition, the deviation modifies the Higgs couplings,
which are stringently constrained by the Higgs data at the LHC
experiments. Within the constraint, both the production cross section
and the branching ratios to the electroweak gauge bosons of a 2~TeV
doublet scalar are found to be extremely small. A higher representation
of $\text{SU}(2)_L$ also suffers from its small production cross section
since it does not couple to the SM quarks directly. For these reasons,
we conclude that it is quite difficult to explain the required event
rate with a new scalar particle, and thus we do not pursue this
possibility in the following discussion.

Another candidate is a spin-one vector boson. Such a particle
naturally appears if a high-energy theory contains additional gauge
symmetry that is spontaneously broken at a certain scale above the
electroweak scale. If the symmetry breaking occurs at the TeV scale,
we expect the masses of the extra gauge bosons to be ${\cal
  O}(1)$~TeV. The gauge bosons are produced at the LHC if quarks are
charged under the extra gauge symmetry. In this paper, we investigate
this possibility. An important caveat here is that such a
TeV-scale vector boson has been severely constrained by the LHC
experiments. The strongest constraint is usually from the Drell-Yan
processes \cite{DrellYanBound, ATLAS:2014wra}; for a 2~TeV vector
boson, its production cross section times the branching ratio in the
lepton final states should be much smaller than 1~fb. This bound makes
it quite difficult to realize a sizable event rate for the diboson
decay channel in most extensions of the SM with new gauge symmetries.

One promising setup to suppress the Drell-Yan processes is given by an
$\text{SU}(2)_L$ singlet heavy charged gauge boson with hypercharge
$\pm 1$, which we denote by $W^\prime$.  Such a $W^\prime$ is
contained in some simple extensions of the SM, such as
$\text{SU}(2)_L\otimes \text{SU}(2)_R \otimes \text{U}(1)_{B-L}$
models \cite{Mohapatra}. This $W^\prime$ couples with right-handed
quarks, as well as right-handed charged leptons and neutrinos. 
If right-handed neutrinos are rather heavy, $W^\prime$ is unable to decay
to leptons and thus evade the Drell-Yan bounds. Since $W^\prime$ couples
to the right-handed quarks, it is sufficiently produced at the
LHC. After the electroweak 
symmetry breaking, the $W^\prime$ bosons mix with the electroweak
gauge bosons, which allows $W^\prime$ to decay into $W$ and $Z$.  In
Sec.~\ref{sec:wprime}, we study whether $W^\prime$ can explain the
observed excess. We find that it is difficult to realize large decay
branch into $WZ$ in a simple version of the $W^\prime$ model, and
therefore the required event rate for the ATLAS diboson excess is not
obtained once the limit from the electroweak precision measurements
is taken into account. Even if this limit is avoided by canceling the
$W^\prime$ contribution to the electroweak precision observables with
other new physics effects, the resonance search in the channel
consisting of a $W$ boson and a Higgs boson ($Wh$)
\cite{Khachatryan:2015bma} severely constrains the $WZ$ decay branch.
Besides, $W'$ generally predicts flavor changing gauge couplings
\cite{Zhang:2007da}, as the $W$ boson does in the SM. We have to assume
these couplings to be flavor-diagonal to evade the strong bounds from 
flavor physics. This gives rise to additional complexity for a concrete
model building in this direction.  

An alternative way is to regard the resonance as a neutral massive
gauge boson $Z^\prime$ which has no coupling to the SM leptons.  This
is the so-called leptophobic $Z^\prime$. 
Such a leptophobic $Z^\prime$ may be realized in the  Grand Unified
Theories (GUTs); if the rank of the GUT group is larger than four, it
includes extra U(1) symmetries, and a certain linear
combination of the U(1) charges could be leptophobic. Especially, a
set of charge assignments inspired by the E$_6$ GUTs has been widely
studied so far in the literature
\cite{E6leptophobic,Babu,E6leptophobic2,Ko,E6leptophobic3}.
The Drell-Yan bounds on this class of models are then
readily avoided because of the leptophobic nature. Again, $Z^\prime$
mixes with the $Z$ boson after the electroweak symmetry breaking, and thus
it has a decay mode into a pair of $W^\pm$. We study the decay
properties of such a $Z^\prime$ using a simplified model in
Sec.~\ref{sec:zprime} to see whether it could explain the ATLAS
diboson signal.

Finally, in Sec.~\ref{sec:summary}, we conclude our discussion and give
some future prospects for probing the scenarios
in the future LHC experiments.

\section{$W^\prime$ model}
\label{sec:wprime}

To begin with, we consider a simplified model for $W^\prime$ to study
whether it explains the ATLAS diboson signal or not. For recent works
on phenomenological studies of $W^\prime$, see
Ref.~\cite{Grojean:2011vu}. As mentioned in the Introduction, we
consider an SU(2)$_L$ singlet vector boson with $+1$ hypercharge as a
candidate for $W^{\prime +}$, since it effectively has no coupling to
the SM leptons and thus avoids the severe Drell-Yan bounds. Such a
vector boson may be attributed to a gauge boson of a non-Abelian gauge
group orthogonal to SU(2)$_L$, like SU(2)$_R$. We may also take up an
SU(2)$_L$ triplet non-hypercharged vector boson, which, for instance,
appears in the $\text{SU}(2)_1\otimes \text{SU}(2)_2 \otimes
\text{U}(1)_Y$ type models \cite{Barger:1980ix}. In this case,
however, couplings of the SU(2)$_L$ triplet vector bosons to the SM
charged leptons are generically allowed, and thus we need an additional
mechanism to suppress these couplings to evade the Drell-Yan
constraints. In this sense, the SU(2)$_L$ singlet vector boson is more
favored, and thus we focus on this candidate in our work.

Let us denote the massive SU(2)$_L$ singlet vector boson by
$\hat{W}^{\prime +}$.  There are scalars charged under the additional SU$(2)$ symmetry,
and they develop nonzero VEVs to cause the SU$(2)$ symmetry breaking.
Then $\hat{W}^{\prime +}$ gains a TeV-scale mass.
We assume that some of the scalars are charged under SU(2)$_L$ as well and  the finite mass mixing between
$\hat{W}^{\prime +}$ and $\hat{W}^+$ in the SM is generated by their
VEVs. The mix is described as
\begin{equation}
\begin{pmatrix}
 W^+ \\ W^{\prime +}
\end{pmatrix}
=
\begin{pmatrix}
 \cos\zeta & \sin\zeta \\ -\sin\zeta & \cos \zeta 
\end{pmatrix}
 \begin{pmatrix}
  \hat{W}^+ \\ \hat{W}^{\prime +} 
 \end{pmatrix}
~,
\end{equation}
where $W^+$ and $W^{\prime +}$ are the mass eigenstates. We
expect the mixing angle $\zeta$ is ${\cal 
O}(v^2/M_{W^\prime}^2)$ where $M_{W^\prime}$ is the mass of $W^\prime$
and $v\simeq 246$~GeV is the Higgs VEV. 
Then, the partial decay width of $W^{\prime +}$ into $W^+$ and $Z$ is
given as follows:
\begin{equation}
 \Gamma(W^{\prime +} \to W^+ Z) \simeq \frac{\alpha_2 \sin^2 2\zeta}{192}
\frac{M_{W^\prime}^5}{M_W^4} ~,
\end{equation}
where $\alpha_2$ is the SU(2)$_L$ gauge coupling and $M_W$ is the mass
of $W$ boson. From this expression, we find that although the partial
decay width is suppressed by the small mixing angle $\zeta$, this
suppression is compensated for by the enhancement factor of
$(M_{W^\prime}/M_W)^4$. This enhancement factor results from the
high-energy behavior of the longitudinal mode of $W^\prime$. Therefore,
we expect a sizable decay branch for the $W^+Z$ channel. The partial
decay width gets increased as the mixing angle becomes large. The size
of the mixing angle is, on the other hand, restricted by the electroweak
precision measurements since it is induced by interactions which
break the custodial symmetry---namely, the bound on the $T$ parameter
\cite{peskin} constrains the mixing angle. The current limit on $\zeta$
is given by $|\zeta|\lesssim 5\times 10^{-4}$ for $M_{W^\prime} = 2$~TeV
\cite{delAguila:2010mx}, which in turn gives an upper limit on
$\Gamma(W^\prime \to W^+ Z)$. We however note that the constraints may
be evaded if there is another contribution to the $T$ parameter which
cancels the effects of the $W$-$W^\prime$ mixing. The actual
realization of this possibility is model-dependent, and we do not pursue
it in this paper. 

The equivalence theorem tells us that the final state gauge bosons in
the $W^{\prime +} \to W^+ Z$ channel could be regarded as Nambu-Goldstone (NG)
bosons, since the longitudinal mode dominates the decay amplitude as we have
just mentioned. Thus, the partial decay width of the channel is
related to that of the decay to $W^+$ and the Higgs boson in the final
state. In fact, we have 
\begin{equation}
 \Gamma(W^{\prime +} \to W^+ h) \simeq \Gamma(W^{\prime +} \to W^+ Z) ~,
\end{equation}
where $h$ is the SM-like Higgs boson. Currently, the CMS collaboration
gives an upper bound on this decay mode \cite{Khachatryan:2015bma} as
$\sigma (pp\to W^{\prime +})\times\text{BR} (W^{\prime +} \to W^+ h)
\lesssim 7$~fb. Thus, through the above equation, this bound also
implies $\sigma (pp\to W^{\prime +})\times\text{BR} (W^{\prime +} \to W^+ Z)
\lesssim 7$~fb, which somewhat conflicts with the ATLAS diboson
anomaly. Since the above relation is a consequence of the equivalence
theorem, this bound is robust and almost model-independent. For this
reason, a $W^\prime$ model (as well as a $Z^\prime$ model as we will see
in the next section) in general predicts smaller number of signals in
the diboson channel than that observed in Ref.~\cite{Aad:2015owa}, once
we consider the limit on the $Wh$ channel.

$\hat{W}^{\prime +}$ carries the $+1$ hypercharge, so that the $\text{SU}(2)_L\otimes \text{U}(1)_Y$ symmetry allows the following couplings the right-handed quarks:
\begin{equation}
 {\cal L}_{W^\prime ud} =
  \frac{g_{ud}}{\sqrt{2}}\overline{u_i}\Slash{\hat{W}}^{\prime +} P_R d_i
+ \text{h.c.} ~,
\label{eq:wprimequarkint}
\end{equation}
where $P_{L/R}\equiv (1\mp \gamma_5)/2$ and $i =1, \,2, \,3$ denotes the
generation index. Here, we assume the coupling constant $g_{ud}$ is
common to all of the generations and ignore flavor non-diagonal parts for
brevity, which are in fact stringently constrained by the measurements of
the flavor observables, such as the $K^0$-$\overline{K}^0$ mass
difference.  At the LHC, 
$W^\prime$ is produced via the interactions in
Eq.~\eqref{eq:wprimequarkint}. For $W^\prime$ with a mass of 2~TeV,
the production cross section at the LHC with the
center-of-mass energy $\sqrt{s}=8$~TeV (LHC8) is evaluated as 
\begin{equation}
 \sigma (pp \to W^{\prime\pm}) \simeq 490 \times g_{ud}^2 ~~[\text{fb}] ~,
\end{equation}
using MadGraph \cite{madgraph}.
After the production, $W^\prime$ mainly decays into the $WZ$, $Wh$, or quark
final states. The partial decay width for the final state containing a
pair of $u_i$ and $\bar{d}_i$ is given by
\begin{equation}
 \Gamma (W^{\prime +} \to u_i \bar{d}_i )= \frac{g_{ud}^2}{16 \pi}
  M_{W^\prime} ~,
\end{equation}
where we neglect the quark masses for brevity.
The branching fraction of this decay mode is severely constrained by the
dijet resonance searches \cite{Khachatryan:2015sja, Aad:2014aqa}. 
Following
Ref.~\cite{Khachatryan:2015sja}, we have $\sigma (pp\to W^\prime)\times
\text{BR}(W^\prime \to q\bar{q}^\prime) \lesssim 100~\text{fb}$ with the
acceptance ${\cal A}\simeq 0.6$ being assumed. The ATLAS collaboration
gives a similar limit on the dijet channel
\cite{Aad:2014aqa}. Currently, the $W^{\prime +} \to t \bar{b}$ decay
mode is less constrained \cite{Aad:2014xea}: $\sigma (pp\to W^{\prime +}
)\times\text{BR} (W^{\prime +} \to t \bar{b}) \lesssim 120$~fb.

\begin{figure}[t]
\centering
\includegraphics[clip, width = 0.6 \textwidth]{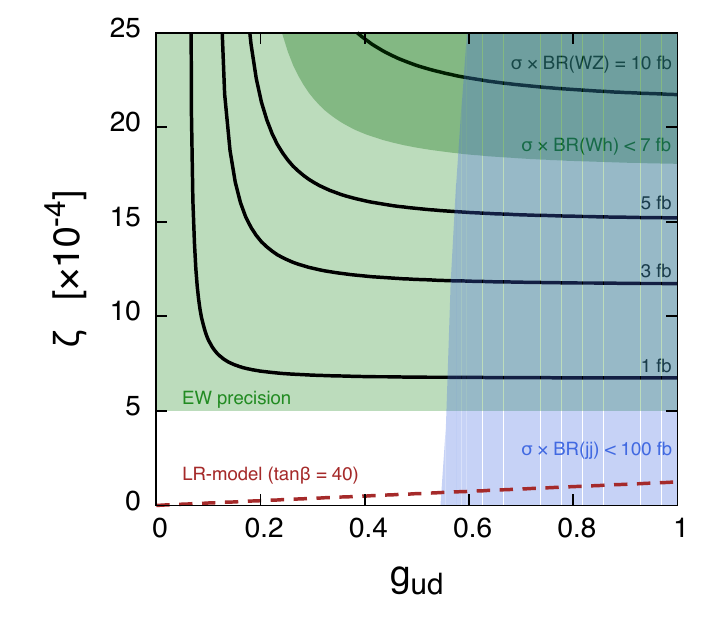}
\caption{$\sigma (pp \to W^\prime) \times \text{BR}(W^\prime \to WZ)$
 (black solid lines). Here we set $M_{W^\prime } = 2$~TeV. Light-green shaded
 region is disfavored by the electroweak precision measurements. Dark-green
 and blue shaded regions are excluded by the limits from the $W^\prime
 \to Wh$ \cite{Khachatryan:2015bma} and the dijet \cite{Khachatryan:2015sja, Aad:2014aqa} channels, respectively. Brown dashed line
 represents the case of the $\text{SU}(2)_L\otimes \text{SU}(2)_R\otimes
 \text{U}(1)_{B-L}$ model
 with $\tan \beta =40$.} 
\label{fig:wlimit}
\end{figure}

Taking the above discussion into account, in Fig.~\ref{fig:wlimit}, we
show a contour plot for $\sigma (pp \to W^\prime) \times
\text{BR}(W^\prime \to WZ)$ on the $g_{ud}$-$\zeta$ plain. The
light-green shaded region is disfavored by the electroweak precision
measurements: $|\zeta| \lesssim 5\times 10^{-4}$. The dark-green and blue shaded
regions are excluded by the limits from the $W^\prime \to Wh$
\cite{Khachatryan:2015bma} and the dijet \cite{Khachatryan:2015sja,
Aad:2014aqa} channels, respectively. This figure clearly shows that it
is difficult to explain the observed diboson resonance with this simplified
$W^\prime $ model once the electroweak precision bound on the
$W$-$W^\prime$ mixing is taken into account. Even if this constraint is
avoided by utilizing other new physics effects to cancel the $W^\prime$
contribution to the electroweak precision observables, the bound on
the $W^\prime \to Wh$ channel restricts $\sigma (pp \to W^\prime) \times
\text{BR}(W^\prime \to WZ)$ to be less than $\sim 7$~fb.

One of the most popular models in which an SU(2)$_L$ singlet $W^\prime$
appears is the so-called left-right (LR) symmetric model based on the
$\text{SU}(2)_L\otimes \text{SU}(2)_R \otimes \text{U}(1)_{B-L}$ gauge
theory. In this model, the $W^\prime$-quark coupling is given by the
SU(2)$_R$ gauge coupling constant $g_R$; $g_{ud} = g_R$. If right-handed
neutrinos in this model are heavier than $W^\prime$, then $W^\prime$
does not decay into the right-handed neutrinos, and thus this model
realizes the setup of the simplified model we have discussed here. 
In this model, the SM Higgs field is embedded into a bi-fundamental
representation of $\text{SU}(2)_L\otimes \text{SU}(2)_R$. Once this
bi-fundamental field acquires the VEV, the $W$-$W^\prime$ mixing is
induced and given by
\begin{equation}
 \tan 2\zeta \simeq 2 \sin 2\beta
  \left(\frac{g_R}{g_L}\right)\frac{M_W^2}{M_{W^\prime}^2} ~,
\end{equation}
where $g_L$ is the SU(2)$_L$ gauge coupling constant and $\tan \beta$ is
the ratio between the diagonal components of the bi-fundamental Higgs
VEV. In Fig.~\ref{fig:wlimit}, we also show the value of $\zeta$ obtained
through this relation as a function of $g_{ud} = g_R$ for $\tan \beta
=40$ in the brown dashed line. This value of $\tan\beta$ is favored to
explain the top and bottom quark masses in this model. It is found that
although the predicted values evade the electroweak precision bound, it
is far below the values required to explain the diboson excess.

Before concluding this subsection, we comment on other possible excesses
reported so far which might also indicate the presence of a $W^\prime$ with a
mass of around 2~TeV. In Ref.~\cite{CMSWh}, the CMS collaboration
reported a  small excess near 1.8~TeV in the searches of $W^\prime$
decaying into $W$ and  Higgs boson in the $l\nu bb$ final
state. However, as we have seen above, this conflicts with another
constraint on the $W^\prime \to Wh$ channel given by the CMS experiment
\cite{Khachatryan:2015bma}. In addition, the 
CMS collaboration announced a possible signal in the searches of
$W^\prime$ decaying into the two electrons and two jets final state
through a right-handed neutrino, whose peak is around $2.1$~TeV with its
significance being $\sim 2.8 \, \sigma$ \cite{Khachatryan:2014dka}. Though
there have been several proposals for $W^\prime$ models that account for the
$2.8 \, \sigma$ excess \cite{Deppisch:2014qpa}, the models in general
predict too small event rates for the diboson channel, and therefore fail
to explain the ATLAS diboson resonance signal.

\section{$Z^\prime$ model}
\label{sec:zprime}

Next, we consider a leptophobic $Z^\prime$. For reviews on $Z^\prime$
models, see Refs.~\cite{Leike:1998wr, Langacker:2008yv}. We regard it
as a gauge boson accompanied by an extra U(1) symmetry, U(1)$^\prime$,
whose mass is generated after the U(1)$^\prime$ gauge symmetry is
spontaneously broken. Suppose that there are two SU(2)$_L$ doublets
and one singlet Higgs bosons $H_u$, $H_d$, and $\Phi$, respectively,
with their $\text{U}(1)^\prime$ charges being $Q_{H_u}^\prime$,
$Q_{H_d}^\prime$, and $Q_\Phi^\prime$. We further assume that $H_u$
only couples to the up-type quarks while $H_d$ couples to the
down-type quarks and charged leptons, just as the minimal supersymmetric SM
(MSSM) and the Type-II two-Higgs-doublet model.  We require
U(1)$^\prime$ to be leptophobic, {\it i.e.}, $Q_{L}^\prime =
Q_{e_R^c}^\prime = 0$, and then this leads to $Q_{H_d}^\prime = 0$.

After these Higgs bosons acquire VEVs, the mass matrix for the
$\text{U}(1)^\prime$ gauge field $\hat{Z}^\prime$ and a linear
combination of the $\text{SU}(2)_L$ and $\text{U}(1)_Y$ gauge fields
($\hat{W}^a$ and $\hat{B}$, respectively),
$\hat{Z} = \cos\theta_W \hat{W}^3 -\sin \theta_W \hat{B}$, is given by
\begin{equation}
 {\cal L}_{\text{mass}} = \frac{1}{2}
(\hat{Z}~\hat{Z}^\prime) 
\begin{pmatrix}
 \hat{M}^2_Z & \Delta M^2 \\ \Delta M^2 & \hat{M}_{Z^\prime}^2
\end{pmatrix}
\begin{pmatrix}
 \hat{Z} \\ \hat{Z}^{\prime}
\end{pmatrix}
~,
\end{equation}
with
\begin{equation}
 \hat{M}^2_Z = \frac{g_Z^2v^2}{4}~, ~~~~~
 \Delta M^2 =-\frac{g_Zg_{Z^\prime}v^2}{2}Q^\prime_{H_u}\sin^2\beta ~,
~~~~~
 \hat{M}_{Z^\prime}^2 = g_{Z^\prime}^2(Q_{H_u}^{\prime 2}\sin^2\beta v^2
+Q^{\prime 2}_\Phi v_\Phi^2) ~.
\end{equation}
Here $ \langle H^0_u\rangle = v \sin \beta/ \sqrt 2 $, $ \langle H^0_d
\rangle = v \cos \beta/ \sqrt 2$, and $ \langle \Phi \rangle = v_\Phi/
\sqrt 2$ are defined.  $g_Z$ is the gauge coupling constant given by
$g_Z \equiv \sqrt{g^{\prime 2}+g^2}$ with $g^\prime$ and $g$ being the
U(1)$_Y$ and SU(2)$_L$ gauge coupling constants, respectively, and
$g_{Z^\prime}$ is the U(1)$^\prime$ gauge coupling constant.  The mass
eigenstates $Z$ and $Z^{\prime}$ are then obtained through the
diagonalization with an orthogonal matrix as
\begin{equation}
 \begin{pmatrix}
  Z \\ Z^{\prime}
 \end{pmatrix}
=
\begin{pmatrix}
 \cos \theta & \sin\theta \\
 -\sin \theta & \cos \theta 
\end{pmatrix}
\begin{pmatrix}
 \hat{Z} \\ \hat{Z}^{\prime}
\end{pmatrix}
~,
\end{equation}
with 
\begin{equation}
 \tan 2\theta = -\frac{2 \Delta M^2}{\hat{M}_{Z^\prime}^2 - \hat{M}^2_Z}
\simeq
4 Q^{\prime}_{H_u}\sin^2\beta
\biggl(\frac{g_{Z^\prime}}{g_Z}\biggr)\frac{M_Z^2}{M_{Z^\prime}^2} 
 ~,
\end{equation}
where $M_Z$ and $M_{Z^\prime}$ are the masses of $Z$ and $Z^\prime$,
respectively. Again, the mixing angle is suppressed by 
a factor of ${M_Z^2}/{M_{Z^\prime}^2}$.   

The couplings of $\hat{Z}^\prime $ to the SM fermions $f$ are given by
\begin{equation}
 {\cal L}_{\text{int}} = g_{Z^\prime}\overline{f} \Slash{\hat{Z}}^\prime
(Q_{f_L}^\prime P_L +Q_{f_R}^\prime P_R) f ~,
\end{equation}
with $Q_{f_L}^\prime$ and $Q_{f_R}^\prime$ the U(1)$^\prime$ charges
of the left- and right-handed components of $f$, respectively. Here again, we
have neglected possible flavor changing effects for simplicity.

Now let us evaluate the partial decay widths of $Z^\prime$. 
For the decay mode into quarks, we have
\begin{equation}
 \Gamma(Z^\prime \to q \bar{q}) = \frac{g_{Z^\prime}^2 N_C}{24\pi}
  M_{Z^\prime} 
\left[
Q_{q_L}^{\prime 2} +Q_{q_R}^{\prime 2}-(Q^{\prime}_{q_L}-Q^{\prime}_{q_R})^2
\frac{m_q^2}{M_{Z^\prime}^2}
\right]\sqrt{1-\frac{4m_q^2}{M_{Z^\prime}^2}}~,
\label{branching-quark}
\end{equation}
where $N_C = 3$ indicates the color factor. 
For the $Z^\prime \to W^+W^-$, on the other hand, we have
\begin{equation}
 \Gamma(Z^\prime \to W^+W^-)
=\frac{g_{Z^\prime}^2}{48\pi} Q^{\prime 2}_{H_u}\sin^4\beta M_{Z^\prime} ~.
\end{equation}
Note that this decay width again remains sizable even though the decay
mode is induced via the $Z$-$Z^\prime$ mixing, since the enhancement
coming from the longitudinal polarization mode compensates for the
suppression factor.  
According to the equivalence theorem,
this decay width becomes equivalent to that of the $Z^\prime \to Z h$
mode in the decoupling limit:
\begin{equation}
 \Gamma(Z^\prime \to Z h) = \Gamma(Z^\prime \to W^+W^-) ~.
\end{equation}
From the above equations, we find that the decay properties of $Z^\prime$ are
determined by the U(1)$^\prime$ charges of quarks and $H_u$, the
U(1)$^\prime$ gauge coupling constant $g_{Z^\prime}$, and $\tan
\beta$. Among them, we can always decrease one degree of freedom via the
redefinition of the U(1)$^\prime$ charge normalization. In what follows,
we normalize the U(1)$^\prime$ charges such that $Q_{H_u}^\prime =
1$. In this case, we have $Q_{u_R}^\prime = 1+ Q^\prime_Q$ and
$Q_{d_R}^\prime = Q_Q^\prime$ with $Q_Q^\prime \equiv Q_{q_L}^\prime$, where the latter equality follows from
$Q^\prime_{H_d} = 0$.

The production cross section of $Z^\prime$ at the LHC8 is estimated as
\cite{Leike:1998wr} 
\begin{equation}
\sigma (pp \to Z^\prime) \simeq 5.2 \times \left ( \frac{2\Gamma(Z^\prime
\to u\overline{u})+\Gamma (Z'\to d \ov{d})}{\rm GeV} \right )   ~[{\rm
fb}]~.  
\end{equation}
Thus, the production cross sections are determined by the quark
U(1)$^\prime$ charges and $g_{Z^\prime}$ once the $Z^\prime$ mass is
fixed.

The production of $Z^\prime$ is stringently limited by the LHC
experiments. For a leptophobic $Z^\prime$, the strong bounds come
from the $Z^\prime \to Zh$ \cite{Khachatryan:2015bma}, dijet
\cite{Khachatryan:2015sja, Aad:2014aqa}, and $t\bar{t}$
\cite{Khachatryan:2015sma} resonance searches. As before, we use $\sigma
(pp\to Z^\prime)\times \text{BR}(Z^\prime \to Zh) \lesssim 7$~fb and
$\sigma (pp\to Z^\prime)\times \text{BR}(Z^\prime \to 
jj) \lesssim 100$~fb with the acceptance ${\cal A}\simeq 0.6$ being
assumed for the latter case. Here, $\text{BR}(Z^\prime
\to jj)$ denotes the branching ratio for decaying into a pair of
quark jets. The limit from the $t\bar{t}$ resonances is found to be the
strongest, as we see below. The limit depends on the width of
$Z^\prime$. For a $2$~TeV $Z^\prime$ with a decay width of 20~GeV, the
bound is given as $\sigma (pp\to Z^\prime)\times \text{BR}(Z^\prime \to
t\bar{t}) < 11$~fb, while if the decay width is 200~GeV, the bound is
relaxed to be 18~fb \cite{Khachatryan:2015sma}.

Similarly to the case of $W^\prime $, the $Z$-$Z^\prime$ mixing angle is
constrained by the electroweak precision measurements. For a $Z^\prime$
model, however, it is not appropriate to merely use the limits on the $T$
parameter to obtain the $Z$-$Z^\prime$ bound, since the presence of
$Z$-$Z^\prime$ mixing modifies the $Z$-boson coupling to the SM fermions
simultaneously. 
In fact, in the case of a leptophobic $Z^\prime$, the
constraints from the electroweak precision measurements are relaxed
because $Z^\prime$ does not couple to electrons \cite{Umeda:1998nq,
Langacker}. It turns out that the present limit on the mixing angle is
given as $\sin \theta \lesssim 0.008$ \cite{Langacker}, which we use in
the following analysis.

\begin{figure}[t]
\centering
\includegraphics[clip, width = 0.6 \textwidth]{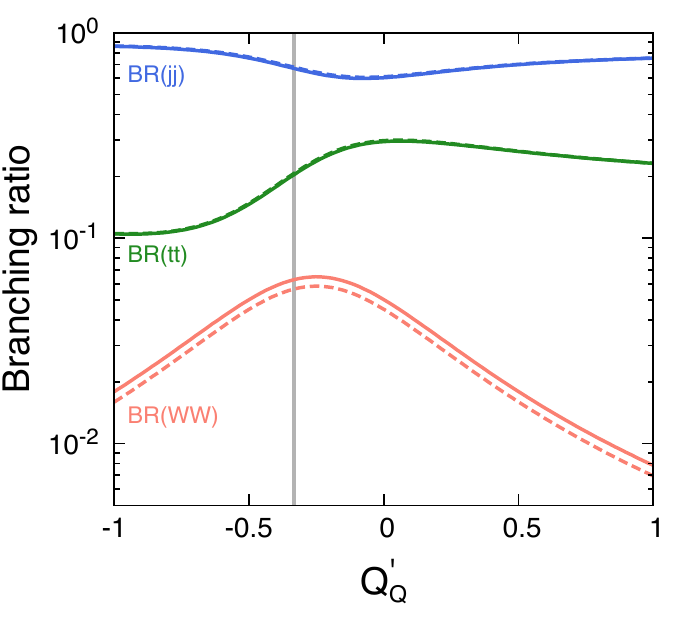}
\caption{Branching ratios of dijet, top quarks, and $WW$ channels
 as functions of $Q^\prime_Q$ from top to bottom. Solid (dashed) lines
 represent the case of $\tan \beta = 40$ (4). We set $M_{Z^\prime} =
 2$~TeV. Vertical gray line corresponds to the charge assignment in the
 E$_6$ inspired leptophobic $Z^\prime$ model mentioned in the text. } 
\label{fig:br}
\end{figure}

\begin{figure}[t]
\centering
\includegraphics[clip, width = 0.6 \textwidth]{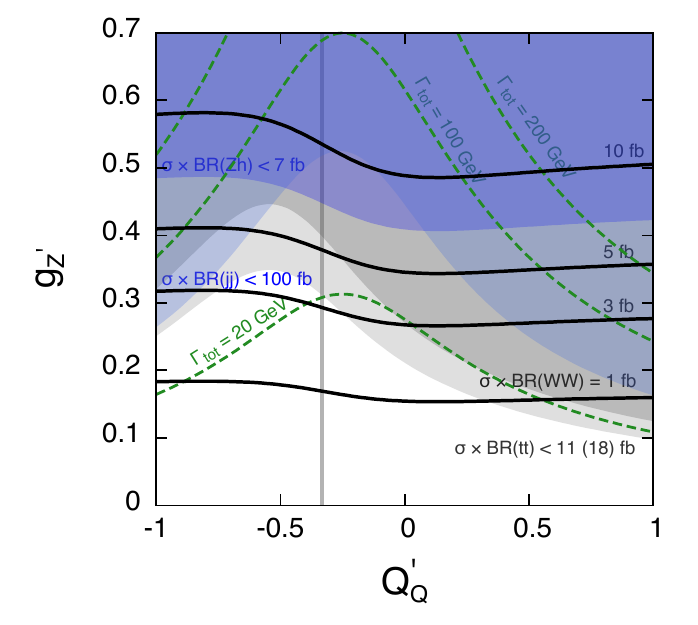}
\caption{Contour for the values of $\sigma (pp\to Z^\prime)\times
 \text{BR}(Z^\prime \to WW)$ in black solid lines. We also show the
 total decay width $\Gamma_{\text{tot}}$ in green dashed lines. Here we
 set $\tan\beta = 40$ and $M_{Z^\prime} = 2$~TeV. Blue and dark-blue
 shaded regions are excluded by the resonance searches in the dijet and
 $Z^\prime \to Zh$ channels, respectively. Dark (light) gray area is
 excluded by the $t\bar{t}$ resonance search if the $Z^\prime$ decay
 width is 200 (20)~GeV. Vertical gray line corresponds to the charge
 assignment in the E$_6$ inspired leptophobic $Z^\prime$ model mentioned
 in the text.}  
\label{fig:zprime}
\end{figure}

In Fig.~\ref{fig:br}, we show branching ratios of the $Z^\prime \to
jj$, $Z^\prime \to t \bar{t}$, and $Z^\prime \to WW$ channels in the
blue, green, and red lines, respectively, as functions of
$Q^\prime_Q$. Here, we set $M_{Z^\prime} = 2$~TeV, and $\tan \beta
=40$ (4) for the solid (dashed) lines. The vertical gray line corresponds to the charge assignment in an
E$_6$ inspired leptophobic $Z^\prime$ model often discussed in the
literature \cite{E6leptophobic,Babu,E6leptophobic2,Ko,E6leptophobic3},
where $Q^\prime_Q = -1/3$, $Q_{u_R}^\prime = 2/3$, $Q_{d_R}^\prime =
-1/3$, and $Q_{\Phi}^\prime = -1$. We
find that the branching fraction for the diboson channel is at most
$\sim 0.05$. The $\tan \beta$ dependence of the branching ratios is
rather small; for instance, if we vary $\tan\beta$ from 40 to 4,
$\text{BR}(Z^\prime\to WW)$ changes by about 10\%. 
Then, in Fig.~\ref{fig:zprime}, we show a contour plot
for the values of $\sigma (pp\to Z^\prime)\times \text{BR}(Z^\prime
\to WW)$ as a function of $g_{Z^\prime}$ and $Q^\prime_Q$. Here, we
set $M_{Z^\prime} = 2$~TeV and $\tan \beta =40$, and the vertical gray
line shows the charge assignments in the E$_6$ inspired leptophobic
$Z^\prime$ model mentioned above. The blue and dark-blue shaded regions are
excluded by the resonance searches in the dijet and $Z^\prime \to Zh$
channels, respectively. The dark (light) gray area is excluded by the
$t\bar{t}$ resonance search if the $Z^\prime$ decay width is 200
(20)~GeV. We also show the total decay width $\Gamma_{\text{tot}}$ in
green dashed lines. 
Contrary to the case of $W^\prime$, the electroweak
precision measurements give no constraint in the parameter region
shown in the figure, since the leptophobic nature of $Z^\prime$
considerably weakens the limit on the $Z$-$Z^\prime$ mixing angle as
discussed above. As can be seen from Fig.~\ref{fig:zprime}, the total
decay width is well below 100~GeV in the allowed parameter region in the
figure. We also find that the $t\bar{t}$ resonance search gives the most
stringent constraint. In particular, if $Q_{Q}^\prime =-1/3$, then
$\sigma (pp\to Z^\prime)\times \text{BR}(Z^\prime \to WW)$ should be
less than about 5~fb, which corresponds to $g_{Z^\prime} \lesssim 0.4$.

Before concluding this section, we discuss an ultraviolet completion of
the simplified leptophobic $Z^\prime$ model. In general, a leptophobic
symmetry causes gauge anomaly, and hence we have to add extra
U(1)$^\prime$-charged chiral fermions so that the contribution of these
extra fermions removes the anomaly. A simple way to find such a set of
extra fermions is to embed the SM particle content into a realization of
an anomaly-free gauge group. Indeed, it turns out that GUTs based on the
supersymmetric (SUSY) E$_6$ gauge group provide a natural framework to realize the leptophobic
$\text{U}(1)^\prime$ symmetry
\cite{E6leptophobic,Babu,E6leptophobic2,Ko,E6leptophobic3}. 
\footnote{For a review of the E$_6$
SUSY GUT models, see, {\it e.g.}, Ref.~\cite{E6review2}. 
}

In SUSY E$_6$ GUTs, all of the MSSM matter fields as well as the
right-handed neutrino superfields are embedded into a ${\bf
27}$-representational superfield in each generation. The ${\bf 27}$
representation also contains new vector-like superfields and a singlet
field with respect to the SM gauge symmetry. A part of these vector-like
fields is identified as the MSSM Higgs fields. 
The rank of E$_6$ group is six, and thus this gauge group
yields two additional U(1) gauge symmetries after the breaking of the
GUT symmetry. It is found to be possible to construct a leptophobic U(1)
charge out of a linear combination of the generators of these extra U(1)
symmetries and $\text{U}(1)_Y$.\footnote{As discussed in
Ref.~\cite{Babu}, a kinetic mixing of the gauge fields associated with
these U(1) symmetries may realize such a desirable linear
combination. However, a concrete realization of the leptophobic
U(1)$^\prime$ model from the E$_6$ GUT requires additional
considerations for subtleties such as gauge coupling unification, the
rapid proton-decay problem, and the structure of the Yukawa
couplings. These possible issues are beyond the scope of the present
paper.} 
The new states in a ${\bf
27}$ are charged under these U(1) symmetries, which make the theory anomaly-free. 
Among the new states, the SM singlet scalar
component may acquire a VEV after the SUSY breaking, which breaks the
leptophobic U(1)$^\prime$ symmetry. If the SUSY breaking occurs around
the TeV scale, we expect the VEV is also ${\cal O}(1)$~TeV. 
Then, the massive gauge boson associated with
the spontaneous breaking 
could be regarded as the 2-TeV leptophobic $Z^\prime$ discussed in this
section. 

As already noted above, the U(1)$^\prime$ charge assignments
for quarks are $Q^\prime_Q = -1/3$, $Q_{u_R}^\prime = 2/3$, and
$Q_{d_R}^\prime = -1/3$. The right-handed neutrinos and $H_u$ have
a unit charge, while the SM singlet scalar component $\Phi$
has $Q^\prime_\Phi = -1$. By definition, the charged leptons have the
zero charge, which results in $Q_{H_d}^\prime = 0$. The masses of the
extra vector-like particles are given by the VEV of $\Phi$. Therefore
these particles also lie around the TeV scale.

This SUSY leptophobic $Z^\prime$ model has several phenomenologically 
interesting features. First, the U(1)$^\prime$ symmetry forbids the 
mass term for Higgsinos, \textit{i.e.}, the $\mu$-term, and it is effectively
induced through the Yukawa coupling between the Higgsinos and the SM
singlet field which breaks the U(1)$^\prime$ symmetry, similar to the
next-to-minimal SUSY SM (NMSSM). As a result, the
Higgsino mass and the $Z^\prime$ mass have the same origin; in
particular, the effective $\mu$-parameter is expected to be around the
TeV scale, which solves the so-called $\mu$-problem. Second, in this
model, there are new tree-level contributions to the Higgs mass: the
$F$-term contribution via the singlet-Higgsino coupling just like the
NMSSM and the $D$-term contribution of the extra U(1)$^\prime$. Taking
into account these contributions as well as the one-loop correction to
the CP-even scalars \cite{Barger:2006dh}, we find that the observed
value of the Higgs mass $\sim 125$~GeV \cite{Aad:2015zhl} is accounted
for with ${\cal
O}(1)$~TeV stops and a small value of $\tan\beta$ when $g_{Z^\prime} =
0.4$--0.5. This should be contrasted with the MSSM prediction; in this case,
if stops have masses of around 1~TeV, the observed Higgs mass is
achieved with a large value of $\tan \beta$, while if $\tan\beta$ is
small then stops in general have masses much larger than ${\cal
O}(1)$~TeV to explain the Higgs mass. Notice that the 2-TeV $Z^\prime$
favors TeV-scale SUSY particles since the SUSY and U(1)$^\prime$
breaking scales are related to each other through the soft mass term for
$\Phi$ in the scalar potential, which triggers the U(1)$^\prime$
breaking. Therefore, this model provides a natural framework for a
light stop scenario without conflicting with the 125~GeV Higgs mass,
which is desirable from the viewpoint of the electroweak fine-tuning
problem. As various particles are predicted to have masses of $\sim
1$~TeV, not only the further investigations of the diboson
events, but also the direct searches of these particles in the LHC
run-II play an important role to test this model.\footnote{We however
note that the decay branching ratios of $Z^\prime$ in this model may be
different from those presented above if some of the additional particles
have masses smaller than $1$~TeV. }

\section{Summary}
\label{sec:summary}

We have considered some extensions of the SM that could explain the
excesses recently reported by the ATLAS collaboration \cite{Aad:2015owa}.
These possible signals are found in the diboson resonance searches with
two fat jets in the final state, and to account for the signals,
production cross sections of $\sim 6$~fb with narrow decay widths are
required. These excesses may be reproduced by an 
extra vector boson with $2$-TeV mass, and we  investigated the $W'$ and
$Z'$ models, especially. $W'$ boson is, for example, predicted by the
additional SU(2)$_R$ gauge symmetry, and decays to not only the SM
fermions but also $W$ and $Z$ bosons through the $W$-$W'$ mixing. There
is a tension between the excess and the bounds from the Drell-Yan
processes, which forces us to forbid the leptonic decay of $W^\prime$ by
making the right-handed neutrinos heavy. 
We also suffer from the tree-level flavor changing couplings of $W'$,
and thus we need to find out a way to forbid the couplings. Besides, the
constraint from the electroweak precision measurements is too severe to
reproduce the excess in the diboson channel. Eventually, we conclude
that it is difficult to interpret the diboson signal as the $W^\prime$
resonance, unless the above difficulties are evaded with additional
conspiracy.

$Z'$ boson is another good candidate for the diboson resonance. It also
appears in various new-physics models; for instance, an extra U(1)$'$
symmetry is predicted by the GUTs based on large gauge groups, in which
the U(1)$^{\prime}$ charges are generally assigned to the SM fermions
and the Higgs field according to their gauge structure. In this case,
the $Z^\prime$ associated with this U(1)$^{\prime}$ symmetry decays
to a pair of the SM fermions, $W^+W^-$ and $Zh$ through the $Z$-$Z'$
mixing generated by the kinetic mixing and mass mixing due to the
nonzero VEV of the U(1)$^\prime$-charged Higgs field. Again, a
leptophobic $Z^\prime$ has an advantage to suppress the Drell-Yan bound.
In Sec.~\ref{sec:zprime}, we investigate such a possibility and
find that there is a sizable parameter region that could
explain a large part of the excess and is still allowed by the current
experimental constraints. We also consider a concrete leptophobic
$Z^\prime$ model inspired by the E$_6$ GUT, and discuss its implication
on the Higgs mass and the SUSY scale. This model predicts the new
vector-like particles and the SUSY particles to have TeV-scale masses,
which are accessible at the next stage of the LHC running. We therefore
expect that the future LHC experiments may not only provide us a deeper
understanding for the ATLAS diboson excess, but also shed light on new
physics behind simplified models discussed in this paper.

Finally we briefly comment on the diboson resonance searches performed
by the CMS collaboration \cite{Khachatryan:2014hpa,
Khachatryan:2014gha}. The CMS collaboration has searched for resonances
decaying into two gauge bosons in hadronic final states
\cite{Khachatryan:2014hpa}, similarly to the ATLAS search
\cite{Aad:2015owa}, as well as in semi-leptonic final states
\cite{Khachatryan:2014gha}. Interestingly, a small excess was found in
both of these searches around 1.8~TeV for the resonance mass, which
might be the same origin as the ATLAS diboson anomaly. If it is not the
case, the semi-leptonic search result \cite{Khachatryan:2014gha} gives a
stringent limit on the 2~TeV excess observed by the ATLAS
collaboration. After all, further searches of the diboson events are
indispensable for confirming or excluding the 2~TeV diboson anomaly,
and are to be done in the near future.

\section*{Acknowledgments}
This work is supported by Grant-in-Aid for Scientific research
from the Ministry of Education, Science, Sports, and Culture (MEXT),
Japan, Grant No. 23104011 (for J.H. and Y.O.). 
The work of J.H. is also supported by World Premier
International Research Center Initiative (WPI Initiative), MEXT,
Japan. The work of N.N. is supported by Research Fellowships
of the Japan Society for the Promotion of Science for Young
Scientists.


\appendix




\begin{thebibliography}{99}

\bibitem{Aad:2015owa} 
  G.~Aad {\it et al.}  [ATLAS Collaboration],
  arXiv:1506.00962 [hep-ex].

\bibitem{Fukano:2015hga} 
  H.~S.~Fukano, M.~Kurachi, S.~Matsuzaki, K.~Terashi and K.~Yamawaki,
  arXiv:1506.03751 [hep-ph].

\bibitem{DrellYanBound}
  G.~Aad {\it et al.}  [ATLAS Collaboration],
  Phys.\ Rev.\ D {\bf 90}, 052005 (2014);
  V.~Khachatryan {\it et al.}  [CMS Collaboration],
  JHEP {\bf 1504}, 025 (2015).

\bibitem{ATLAS:2014wra} 
  G.~Aad {\it et al.}  [ATLAS Collaboration],
  JHEP {\bf 1409}, 037 (2014);
  V.~Khachatryan {\it et al.}  [CMS Collaboration],
  Phys.\ Rev.\ D {\bf 91}, 092005 (2015).

\bibitem{Mohapatra} 
  R.~N.~Mohapatra and J.~C.~Pati,
  Phys.\ Rev.\ D {\bf 11}, 566 (1975);
  Phys.\ Rev.\ D {\bf 11}, 2558 (1975);
  G.~Senjanovic and R.~N.~Mohapatra,
  Phys.\ Rev.\ D {\bf 12}, 1502 (1975).


\bibitem{Khachatryan:2015bma} 
  V.~Khachatryan {\it et al.}  [CMS Collaboration],
  arXiv:1506.01443 [hep-ex].



\bibitem{Zhang:2007da} 
  Y.~Zhang, H.~An, X.~Ji and R.~N.~Mohapatra,
  Nucl.\ Phys.\ B {\bf 802}, 247 (2008);
  A.~Maiezza, M.~Nemevsek, F.~Nesti and G.~Senjanovic,
  Phys.\ Rev.\ D {\bf 82}, 055022 (2010);
  D.~Guadagnoli and R.~N.~Mohapatra,
  Phys.\ Lett.\ B {\bf 694}, 386 (2011);
  S.~Bertolini, A.~Maiezza and F.~Nesti,
  Phys.\ Rev.\ D {\bf 89}, 095028 (2014).


\bibitem{E6leptophobic}
  F.~del Aguila, G.~A.~Blair, M.~Daniel and G.~G.~Ross,
  Nucl.\ Phys.\ B {\bf 283}, 50 (1987);
    V.~D.~Barger, K.~m.~Cheung and P.~Langacker,
  Phys.\ Lett.\ B {\bf 381}, 226 (1996);
  J.~L.~Rosner,
  Phys.\ Lett.\  B {\bf 387}, 113 (1996);
  K.~Leroux and D.~London,
  Phys.\ Lett.\ B {\bf 526}, 97 (2002).



\bibitem{Babu}
  K.~S.~Babu, C.~F.~Kolda and J.~March-Russell,
  Phys.\ Rev.\  D {\bf 54}, 4635 (1996);
  T.~G.~Rizzo,
  Phys.\ Rev.\ D {\bf 59}, 015020 (1998);
  Phys.\ Rev.\ D {\bf 85}, 055010 (2012).

\bibitem{E6leptophobic2} 
  M.~R.~Buckley, D.~Hooper and J.~L.~Rosner,
  Phys.\ Lett.\ B {\bf 703}, 343 (2011).

  \bibitem{Ko} 
  P.~Ko, Y.~Omura and C.~Yu,
  JHEP {\bf 1506}, 034 (2015).



\bibitem{E6leptophobic3} 
  C.~W.~Chiang, T.~Nomura and K.~Yagyu,
  JHEP {\bf 1405}, 106 (2014).



\bibitem{Grojean:2011vu} 
  C.~Grojean, E.~Salvioni and R.~Torre,
  JHEP {\bf 1107}, 002 (2011);
  D.~Pappadopulo, A.~Thamm, R.~Torre and A.~Wulzer,
  JHEP {\bf 1409}, 060 (2014);
  N.~Vignaroli,
  Phys.\ Rev.\ D {\bf 89}, 095027 (2014).

\bibitem{Barger:1980ix} 
  V.~D.~Barger, W.~Y.~Keung and E.~Ma,
  Phys.\ Rev.\ D {\bf 22}, 727 (1980).

\bibitem{peskin}
  M.~E.~Peskin and T.~Takeuchi,
  Phys.\ Rev.\ D {\bf 46}, 381 (1992).

\bibitem{delAguila:2010mx} 
  F.~del Aguila, J.~de Blas and M.~Perez-Victoria,
  JHEP {\bf 1009}, 033 (2010).




\bibitem{madgraph}
  F.~Maltoni and T.~Stelzer,
  JHEP {\bf 0302}, 027 (2003).


\bibitem{Khachatryan:2015sja} 
  V.~Khachatryan {\it et al.}  [CMS Collaboration],
  Phys.\ Rev.\ D {\bf 91}, 052009 (2015);
  S.~Chatrchyan {\it et al.}  [CMS Collaboration],
  Phys.\ Rev.\ D {\bf 87}, 114015 (2013).

\bibitem{Aad:2014aqa} 
  G.~Aad {\it et al.}  [ATLAS Collaboration],
  Phys.\ Rev.\ D {\bf 91}, 052007 (2015).

\bibitem{Aad:2014xea} 
  G.~Aad {\it et al.}  [ATLAS Collaboration],
  Phys.\ Lett.\ B {\bf 743}, 235 (2015).

\bibitem{CMSWh}
CMS Collaboration, CMS-PAS-EXO-14-010.

\bibitem{Khachatryan:2014dka} 
  V.~Khachatryan {\it et al.}  [CMS Collaboration],
  Eur.\ Phys.\ J.\ C {\bf 74}, 3149 (2014).

\bibitem{Deppisch:2014qpa} 
  F.~F.~Deppisch, T.~E.~Gonzalo, S.~Patra, N.~Sahu and U.~Sarkar,
  Phys.\ Rev.\ D {\bf 90}, 053014 (2014);
  M.~Heikinheimo, M.~Raidal and C.~Spethmann,
  Eur.\ Phys.\ J.\ C {\bf 74}, 3107 (2014);
  J.~A.~Aguilar-Saavedra and F.~R.~Joaquim,
  Phys.\ Rev.\ D {\bf 90}, 115010 (2014);
  J.~Gluza and T.~Jeli\'nski,
  Phys.\ Lett.\ B {\bf 748}, 125 (2015).


\bibitem{Leike:1998wr} 
  A.~Leike,
  Phys.\ Rept.\  {\bf 317}, 143 (1999).

\bibitem{Langacker:2008yv} 
  P.~Langacker,
  Rev.\ Mod.\ Phys.\  {\bf 81}, 1199 (2009).

\bibitem{Khachatryan:2015sma} 
  V.~Khachatryan {\it et al.}  [CMS Collaboration],
  arXiv:1506.03062 [hep-ex];
  The ATLAS collaboration,
  ATLAS-CONF-2015-009, ATLAS-COM-CONF-2015-010.

\bibitem{Umeda:1998nq} 
  Y.~Umeda, G.~C.~Cho and K.~Hagiwara,
  Phys.\ Rev.\ D {\bf 58}, 115008 (1998);
  G.~C.~Cho, K.~Hagiwara and Y.~Umeda,
  Nucl.\ Phys.\ B {\bf 531}, 65 (1998)
  [Nucl.\ Phys.\ B {\bf 555}, 651 (1999)].

\bibitem{Langacker} 
  J.~Erler, P.~Langacker, S.~Munir and E.~Rojas,
  JHEP {\bf 0908}, 017 (2009).







\bibitem{E6review2} 
 S.~F.~King, S.~Moretti and R.~Nevzorov,
  Phys.\ Rev.\ D {\bf 73}, 035009 (2006).







\bibitem{Barger:2006dh} 
  V.~Barger, P.~Langacker, H.~S.~Lee and G.~Shaughnessy,
  Phys.\ Rev.\ D {\bf 73}, 115010 (2006).

 
\bibitem{Aad:2015zhl} 
  G.~Aad {\it et al.}  [ATLAS and CMS Collaborations],
  Phys.\ Rev.\ Lett.\  {\bf 114}, 191803 (2015).
  

\bibitem{Khachatryan:2014hpa} 
  V.~Khachatryan {\it et al.} [CMS Collaboration],
  JHEP {\bf 1408}, 173 (2014).

\bibitem{Khachatryan:2014gha} 
  V.~Khachatryan {\it et al.} [CMS Collaboration],
  JHEP {\bf 1408}, 174 (2014).


\end{thebibliography}
\end{document}